\documentclass[letterpaper,11pt]{article}
\usepackage{geometry}         

\usepackage{algorithm,algorithmic,bbold}

\usepackage{graphicx}
\DeclareGraphicsExtensions{.pdf,.jpeg,.png}
\usepackage{amsmath}

\DeclareMathOperator*{\argmin}{arg\,min}
\newcommand{\ks}{{$k$-space}}
\newcommand{\beq}{\begin{equation}}
\newcommand{\eeq}{\end{equation}}
\newcommand{\tp}{\mathrm{TP}}
\newcommand{\fp}{\mathrm{FP}} 
\newcommand{\tn}{\mathrm{TN}}
\newcommand{\fn}{\mathrm{FN}}

\begin{document}
\renewcommand{\thefootnote}{\arabic{footnote}}

\title{\large\bf Robust and Automated Method for Spike Detection and Removal\\ in Magnetic Resonance Imaging}

\author{David~S.~Smith\footnote{Institute of Imaging Science and Department of
Radiology and Radiological Sciences, Vanderbilt University Medical Center,
Nashville, TN}~\footnote{Corresponding author: \texttt{david.smith@vumc.org}}, Joel Kullberg, Johan Berglund, Malcolm J. Avison, and E.~Brian~Welch}

\maketitle

\begin{abstract}

Radio frequency (RF) spike noise is a common source of exogenous
image corruption in MRI. Spikes occur as point-like disturbances
of $k$-space that lead to global sinusoidal intensity errors in the image
domain. Depending on the amplitude of the disturbances and their
locations in $k$-space, the effect of a spike can be significant, often
ruining the reconstructed images. Here we present both a spike
detection method and a related data correction method for automatic
correction of RF spike noise. To detect spikes, we found the $k$-space
points that have the most significant effect on the total variation
of the image.  To replace the spikes, we used a compressed sensing
reconstruction in which only the points thought to be corrupted are
unconstrained.  We demonstrated our technique in two cases: (1) in vivo gradient echo brain data with artificially corrupted points and (2) actual, complex scanner data 
from a whole-body fat-water imaging gradient echo protocol corrupted
by spikes at uncertain locations.  Our method allowed
near-perfect detection and correction with no human intervention. We calculated Matthews correlation coefficients and sensitivities above 0.95 for a maximum of 0.78\% corruption in synthetically corrupted in vivo brain data.  We also found specificities above 0.9994.

\end{abstract}

\section{Introduction}

Radio frequency (RF) spike noise in magnetic
resonance imaging (MRI) typically manifests as bursts of
high-amplitude corruptions in the Fourier domain (\ks) that lead to
Moir\'e patterns in the reconstructed image.  Spikes originate from
brief disruptions in the electromagnetic field near the receive coil
during periods when the receiver channel is open during an exam.
Common sources are static discharge in clothes, mechanical stress and
vibration in the scanner, receive hardware failures, and leaks in the
RF shield that permit external RF interference.

Traditional spike detection techniques rely on thresholding based on
the RF receive signal amplitude \cite{Staemmler1986,Foo1994},
statistical analyses of data \cite{Zhang2001,Chavez2009}, or window
filters \cite{Kao2000}.  With the exception of \cite{Kao2000}, all
detection has been performed entirely in $k$-space. Detected spikes
are then typically replaced by zeros or by local interpolation of
\ks~neighbors \cite{Staemmler1986,Foo1994,Chavez2009}.  In one notable
case \cite{Kao2000}, an analytic solution to the missing data was
used.

The recent application of compressed sensing (CS)
\cite{Candes2006,Donoho2006a} to MRI has allowed the acquisition of MRI data
with a significant relaxation of sampling requirements
\cite{Lustig2008}. In many cases, CS techniques can allow for
reconstruction of full MRI data sets from just 25\% or less of the
full data set, as would historically be defined by the Nyquist-Shannon
sampling theorem. The ephemeral nature of RF spikes, which leads to local corruption of the \ks~data, suggests that a CS
reconstruction of the corrupted data should be extremely accurate.

We were initially motivated to find an improved spike detection and
removal algorithm because we experienced severely corrupted data in a
longitudinal fat-water imaging study. The data were taken with a
faulty RF room shield, and the built-in commercial software was unable
to correct the severely corrupted data on a majority of slices.  In
this work, we present our resulting improved and robust methods for
both spike detection and removal that require no user intervention or
prior information about the acquisition of the data.  This method
can also be run retrospectively when post processing already acquired
data.

In Section 2, we describe a new method for spike detection and
removal.  In Section 3, we show the results of this method on both
in vivo brain data with synthesized spikes and 
severely corrupted actual whole-body gradient echo data, acquired with a
damaged RF room shield beneath the patient support and containing an
unknown number of spikes arising from RF leakage secondary to loss of RF shield integrity.  In Section 4, we discuss the practical uses
for our methods and possible improvements.

\section{Materials and Methods}\label{sec:methods}

\subsection{Spike Detection Method}

During collection of the data at an MRI scanner, the imaging volume is
encoded by variable spatial frequencies that are then interpreted as
the coefficients of the discrete Fourier transform (DFT) of the
target.  Perturbing a single Fourier coefficient in \ks~will lead to
global sinusoidal intensity variations in the image.  In one sense,
the coefficients of the DFT are the coefficients which minimize the
approximation error of the imaging target.  Since anatomical
structures are typically smooth objects on a flat (signal free)
background, the coefficients that cause the most ``rippling'' in the
image should be the least accurate.  Furthermore, since all coefficients
are corrupted at some level by noise, we would like to distinguish a
point of diminishing returns, beyond which we are replacing noisy, but
probably valid, data.  Optimal image improvement should result when
all data corrupted by RF spikes are replaced and all valid points are
retained. In practice, however, this may not necessarily be possible, so we must plan to strike a balance between possibly missing spikes and not correcting all image artifacts or possibly discarding valid data and losing the information that it contained.

We define the collected data, or $k$-space, of a complex-valued image
$\boldsymbol{x}$ as $\boldsymbol{d} = \mathcal{F}\boldsymbol{x}$,
where $\mathcal{F}$ is the DFT operator. To begin spike detection, each element
of $\boldsymbol{d}$ was zeroed in turn, and a corresponding
modified magnitude image $\boldsymbol{x}^k$ was created with an inverse DFT: \beq \boldsymbol{x}^k =|
\mathcal{F}^{-1} \boldsymbol{I}^k \boldsymbol{d}|,\eeq where
$\boldsymbol{I}^k$ is the identity matrix with the $k$-th element
along the diagonal set to zero.  In theory, if the datum $d_k$ deleted was valid, the image quality of $\boldsymbol{x}^k$ should decrease; if $d_k$ had been corrupted, the image quality should improve. Since the total variation (TV) is very sensitive to alterations in the Fourier domain, we hypothesized that the TV should separate the corrupted data from the valid data.

 A vector of aggregated TV values $\boldsymbol{t}$ was then constructed, where the $k$-th element of $\boldsymbol{t}$ was computed as the TV of the derived image $\boldsymbol{x}^k$:  \beq t_k= \sum_{s=1}^N
\left\|\nabla_s \boldsymbol{x}^k\right\|_1,\eeq where $N$ is the total number of data points, $\nabla_s$ is
the forward difference along the dimension $s$ (e.g. $s=\{1,2,3\}$ for
a 3-D protocol) and $\|\cdot\|_1$ is the $\ell_1$-norm, defined for a
vector $\boldsymbol{v}$ as $ \|\boldsymbol{v}\|_1 = \sum_i
|v_i|.$

Next, the upper half of $\boldsymbol{t}$ was assumed to be valid and were discarded (these points had the least effect on image total variation).  The remaining $\boldsymbol{t}$ values were retained and normalized to lie between 0 and 1.  Then a threshold $\theta$ was selected using Otsu's method \cite{Smith1979} (see, e.g., Fig.~\ref{tv}).  Since the penalty in terms of image quality is more severe when corrupted data is missed than when clean data is thought to be corrupted, we increased  the cutoff between corrupted and clean data to $\sqrt{\theta}$ (recall that $\boldsymbol{t}$ has been normalized to be between zero and one). This provided a more aggressive selection of spiked data and will be shown below to be the most robustly accurate method. 

All $\boldsymbol{t} < \sqrt{\theta}$   were assumed to be spikes. 
The suspected spikes were
then discarded to create a data constraint mask $\boldsymbol{M}$ for
the reconstruction: $ \boldsymbol{M} = \prod_{\hat{k}}
\boldsymbol{I}^{\hat{k}},$ where $\hat{k} \in  
\{k : t_k < \sqrt{\theta}\}$. The spike detection algorithm is listed in
Algorithm 1.\begin{algorithm}
\caption{:: Spike Detection}
\begin{algorithmic} 
\REQUIRE input data $\boldsymbol{d}$
\FORALL{data points $k$}
\STATE $\boldsymbol{g} \leftarrow \boldsymbol{d}$
\STATE ${g}_k \leftarrow 0$
\STATE ${t}_k \leftarrow \mathrm{TV}(\mathcal{F}^{-1} \boldsymbol{g})$
\ENDFOR
\STATE Discard upper half of $\boldsymbol{t}$.
\STATE Normalize remaining $\boldsymbol{t}$ to lie on $[0,1]$ interval.
\STATE Calculate Otsu threshold $\theta$ for  $\boldsymbol{t}$.
\STATE All $\boldsymbol{t} < \sqrt{\theta}$ are labeled as assumed spikes.
\RETURN
\end{algorithmic}
\end{algorithm}

\subsection{Collection of Real Data Free of Spikes}

First, a gradient echo data set was used that was free of spikes to use as a baseline for adding synthetic spikes. This gave us a benchmark for accuracy testing of the algorithm. This data was acquired on a Philips 7T scanner with an axial field of view (FOV) of 256 mm $\times$ 256 mm and an in-plane voxel size of 1 mm $\times$ 1 mm. The slice thickness was 2.5 mm. The flip angle was 35 deg, and no partial Fourier or parallel imaging was used. The TE1/$\Delta$TE/TR was 1.74/2.3/200 ms. Total scan time for the acquisition of 32 echoes was 205 s.  
 A reconstructed matrix size of 256$^2$ was produced by gridding 512 radial profiles. This gridded Cartesian data was inverse transformed to produce a 256 $\times$ 256 Cartesian image.  This final, complex image was our baseline ``clean'' image that synthetic spikes were then added to for the synthetic spike corruption experiment.

\subsection{Collection of Real Data Corrupted by Spikes} 
  
As a real-world application, we applied the algorithm to gradient 
echo data from an \emph{in vivo} fat-water MRI protocol.  For this
acquisition, the subject entered a
dual-transmit Philips 3.0 T Achieva TX scanner (Philips Healthcare,
Best, The Netherlands) feet-first in a supine position with arms
extended above the head. The integrated quadrature body coil was used
for both transmit and receive. A multi-station protocol with 20 table
positions was used to acquire whole-body data. Each of the 20 stacks
consisted of a multi-slice, multiple gradient echo acquisition with 12
contiguous 8 mm slices (240 slices total).  Other acquisition details
include: TR/TE1/TE2/TE3 [ms] = 75/1.34/2.87/4.40; flip angle
20$^\circ$; water fat shift = 0.325 pixels (BW = 1335.5 Hz/pixel); axial FOV
= 500 mm $\times$ 390 mm, acquired matrix size = 252 $\times$ 195;
acquired voxel size = 2 mm $\times$ 2 mm $\times$ 8 mm. First-order
shimming was performed for each slice stack. The total duration of
data acquisition was 4 minutes and 16 seconds. Approximately 5 minutes
of additional time was needed for table movement, preparation phases
at each table position, and for breath holding pauses. Breath holding
was performed for table positions covering the waist to the shoulders.

Reconstruction of water and fat images from the acquired multi-echo
data was performed using a generalized three-point Dixon approach
\cite{Berglund2010} in which two solutions were found analytically in
each voxel. Fat and water signal components were found by least
squares fitting after the true solution was identified by imposing
spatial smoothness in a 3D multi-seeded region growing scheme with a
dynamic path that allowed low confidence regions to be solved after
high confidence regions were solved.

\subsection{Reconstruction of Corrected Images}

Images cleaned of spikes were then reconstructed using an
unconstrained TV-regularized compressed sensing reconstruction that
iteratively solved the following optimization problem: \beq
\boldsymbol{x}_\mathrm{clean} = \argmin_{\boldsymbol{x}}
\mathrm{TV}(\boldsymbol{x}) + \frac{\lambda}{2}\|\boldsymbol{M}
\mathcal{F} \boldsymbol{x} - \boldsymbol{M} \boldsymbol{d}\|_2^2,\label{reconeq}\eeq
where $\lambda$ is a scalar weighting factor that controls the balance
between promoting sparsity of the gradient image and fidelity to the
measured non-spiked data.  (The total variation is the $\ell_1$-norm of the gradient image.) For the results here, $\lambda$ was chosen to be high enough that excessive smoothing of the images was avoided; e.g., $\lambda = 50$ for unit-normed data.  Since we only tested this on Cartesian
data, we were able to use a fast split Bregman solver
\cite{Goldstein2009} to solve the reconstruction in Eq. \ref{reconeq}.  The reconstruction code
was adapted to run on a compute server with graphics processing unit
(GPU) acceleration using Jacket 2.2 (AccelerEyes, Atlanta, GA) and
MATLAB R2012a (MathWorks, Natick, MA).

\section{Results}\label{sec:results}

\subsection{Application to Data Synthetically Corrupted}

We first applied the spike detection algorithm to a clean gradient echo data set
of a healthy brain.  This data set was originally spike free to the best of our knowledge.  To simulate the effect of spiking on the data set in a situation in which the positions of the spikes were known, we added varying numbers of synthetically generated spikes to this data.
Random locations in the $k$-space data were replaced with spikes with magnitudes equal to the DC coefficient and a phases randomly chosen between 0 and $2\pi$. 
Real spikes can, and probably will, have varying amplitudes, but choosing to fix the magnitude allowed us to concentrate on the detection accuracy in controlled conditions.  For example, spikes of smaller magnitude would have been harder to detect, but they also would have a smaller influence on image quality, so the penalty for not detecting them would be smaller.  Rather than explore the full problem space of possible spike forms, we chose a simple case for this test.  The detection and correction required 7.4 s per case on a GPU-accelerated workstation.

\begin{figure*}[h]
\centering
\includegraphics[width=\textwidth]{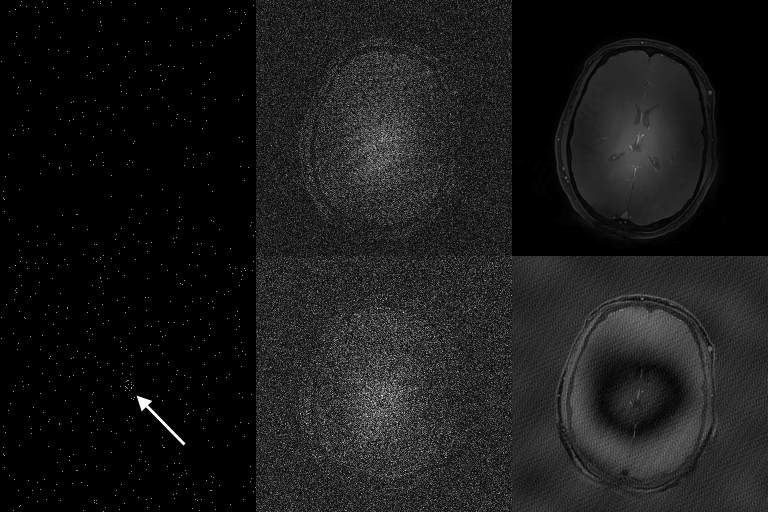}

\caption{Two examples of 7T brain gradient echo images with synthetic spikes added. The first (top row) is the case with the largest number of spikes (243) that still achieved a perfect reconstruction. The second case (bottom row) was the most severely corrupted case, with 404 spikes added.  In this case, the algorithm labeled several points at low spatial frequencies as spikes (arrow), and the severe image artifact can be seen to be a very low spatial frequency feature. This suggests that some of the points deleted and replaced were in fact valid and that they were incorrectly replaced.}
\label{montage}
\end{figure*}

The spike locations, corrupted images, and cleaned images are shown in Fig. \ref{montage} for two cases: (1) the case with the largest number of spikes that still achieved perfect recovery, and (2) the case with the worst corruption in which a low-frequency datum was corrupted and our algorithm failed to recover it correctly.
While Case 2 appears to be a failure of the algorithm, it was notably a case with 404 spikes.  Case 1 was corrupted by 243 spikes and still achieved perfect recovery.  One can see by comparing the middle images of each row in Fig. \ref{montage} to the rightmost images how stark the difference is between the corrupted and recovered images and how dramatically the algorithm cleans up the corrupted images.

\begin{figure}[h]
\centering
\includegraphics[width=3in]{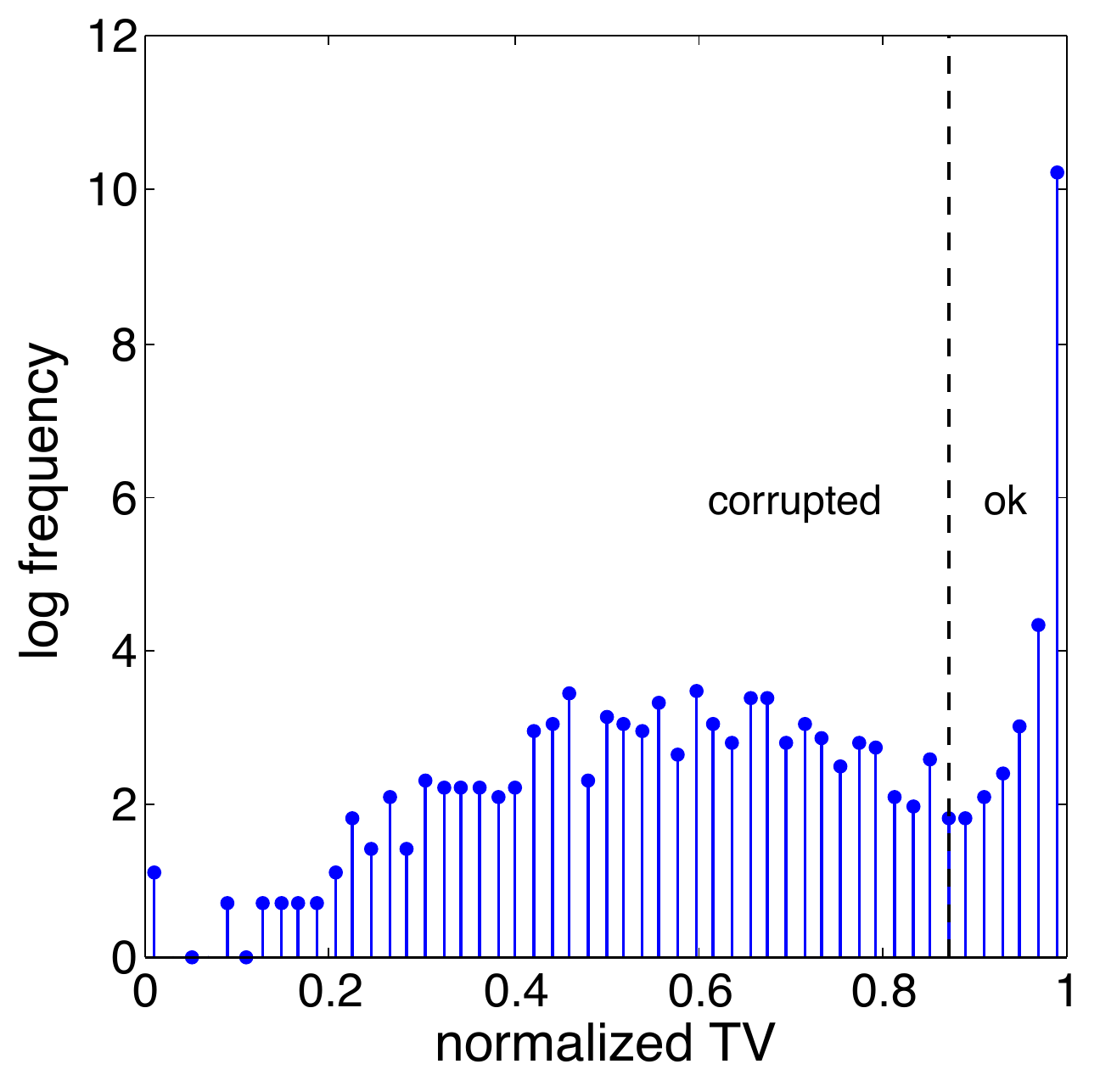}

\caption{Total variation histogram produced when synthetic spikes were added to 7T gradient echo brain data. Only the lower half of TV values are retained, and then Otsu's method is performed to determine a cutoff below which points are considered to be spike corrupted. }

\label{tv}
\end{figure}

A demonstration of the way in which the algorithm separates spikes from genuine data is shown in Fig. \ref{tv}.  Here the lower half of the TV has been normalized and plotted in a histogram. The vertical dashed line is placed at the cutoff $\sqrt{\theta}$.  Data that, when zeroed out, produce TV values less than the cutoff---to the left of the dashed line---are suspected spikes. Points to the right of the line are assumed to be correct.  The histogram shows a bimodal distribution of TV values: a broad distribution at low values and a very narrow, highly peaked distribution near unity.  Otsu's method chooses a TV cutoff value that minimizes the variance between these two distributions, but the histogram suggests that both distributions have overlapping tails, and hence the detection algorithm will in case with many spikes incorrectly label some valid points as spikes and spikes as valid. The goal is to minimize the number of such misclassifications, and Otsu's method is well suited to that goal.

\begin{figure}[h]
\centering
\includegraphics[width=2.8in]{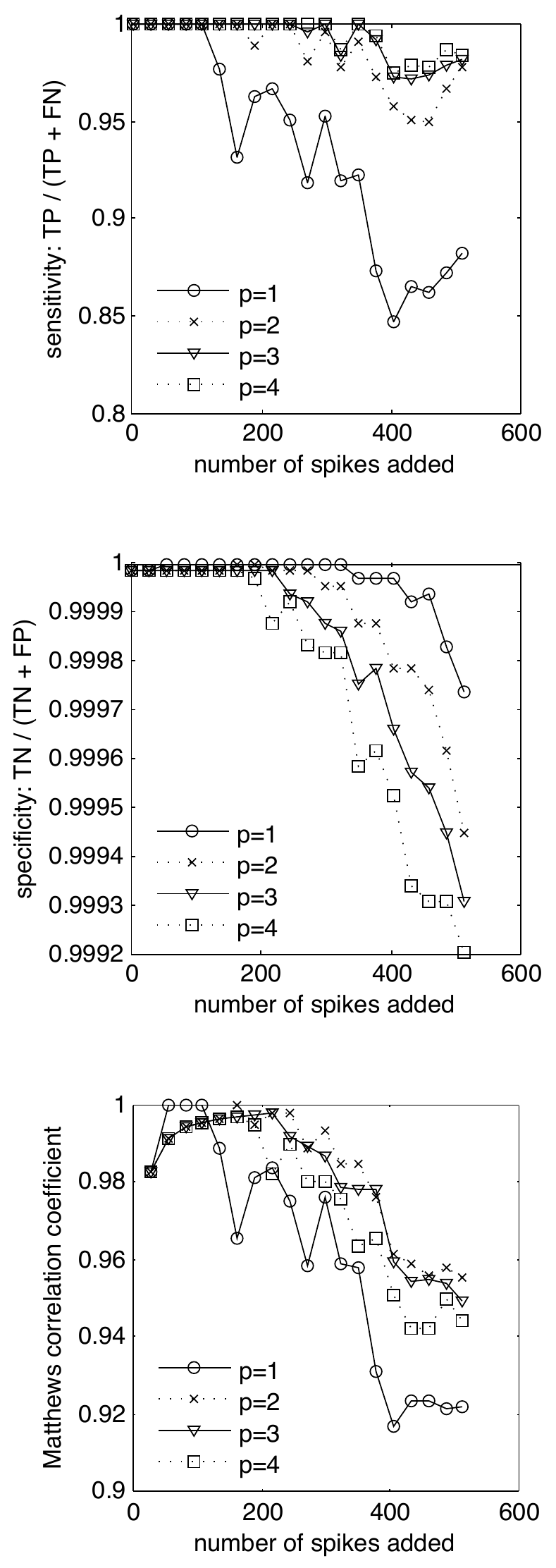}
\caption{Numerical experiment showing the robustness of the spike detection accuracy to the number of spikes added in the synthetically spiked brain example and to adjustments to the threshold given by Otsu's method.  Each curve with points corresponds to raising the Otsu's threshold to the specified fractional power ($1/p$, where $p \in \{1,2,3,4\}$), and the solid black line is perfect agreement with the number of spikes added (i.e. 100\% detection accuracy).  Taking the square root of the Otsu threshold ($p=2$) seems to give the most robust accuracy as the number of spikes increases. }
\label{analysis}
\end{figure}

To quantitatively evaluate the performance of our spike detection algorithm, we measured the sensitivity, specificity, and Matthews correlation coefficient of the algorithm as the number of synthetically generated spikes increased using four different exponents for the Otsu's method cutoff: $1/p$, where $p=\{1,2,3,4\}$.  The results are shown in Fig. \ref{analysis}.  The sensitivity and specificity can be written as 
$\tp/(\tp+\fn)$
and
$\tn/(\tn+\fp),$ respectively, 
where $\tp$ is the number of true positives, $\tn$ is the number of true negatives, $\fp$ is the number of false positives, and $\fn$ is the number of false negatives.

Since spike corruption tends to affect only a small fraction of the  data set, the usual measure of detection accuracy, defined as $(\tp+\tn)/\mathrm{total\ points}$, is unreliable.  Instead we choose to  present the Matthews correlation coefficient (MCC) of the classifications produced by the algorithm (Fig. \ref{analysis}c). The MCC can be thought of as a correlation coefficient between the measured and predicted classification of each point (corrupted or ok). The MCC can be calculated as $$\mathrm{MCC} = \frac{\tp \times \tn - \fp \times \fn}{\sqrt{(\tp+\fp)(\tp+\fn)(\tn+\fp)(\tn+\fn)}}.$$

We found that all three measures decline as more spikes were added, presumably because the number of real points that randomly had low TV values increased.  Sensitivities were very uniformly high ($>$ 0.95) for $p > 1$.  The worst sensitivity was 0.85 for $p=1$ at 404 spikes added.

The specificity of the technique is uniformly very high ($>$ 0.999) no matter how many spikes were added or the exact value of $p$.  This is because the vast majority of data points were never selected, and even a few false positives are insignificant compared to the total number of data points.

Finally, the MCC was above 0.9 for all $p$ up to the maximum number of spikes, but $p=2$ was the highest at moderate to high spike corruption levels.  For very small numbers of spikes, $p=1$ produced a higher MCC, but all $p$ produced high coefficients ($>0.98$) at low corruption levels.   Since correcting data sets with higher levels of spike correction is a more difficult problem, we believe that $p=2$ is the best choice.  This has suggested the choice of $\theta^{1/2}$ for the cutoff, which we adopt throughout.

\subsection{Application to Data Corrupted by Real Spikes}

Next, to test Algorithm 1 under real-world conditions with spikes at 
uncertain locations, we applied it to
\emph{in vivo} whole-body fat-water imaging gradient echo data 
acquired with a damaged RF room shield. RF interference had
severely corrupted the data on almost every slice.  These data were
corrected using Algorithm 1 and then passed to a fat-water separation
algorithm \cite{Berglund2010}  to compare the effects of the data scrubbing on
derived quantitative parameters.  Data from each echo time were processed separately. The total processing time necessary to 
correct all slices and echoes of this data set was approximately 30 min 
on a GPU-accelerated workstation using MATLAB R2012a 
(MathWorks, Natick, MA).

\begin{figure}[h]
\centering
\includegraphics[width=3.5in]{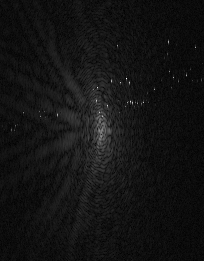}
\caption{Example of spike corruption in $k$-space from the fat-water imaging data set acquired in the \emph{in vivo} experiment.  The bright points that don't follow the general pattern of data are almost certainly corrupted.  Low intensity features have been scaled up to aid visualization by taking the square root of the magnitude of the coefficients.  This is the same data used for the uncorrected images in Fig. \ref{real}.}
\label{realdata}
\end{figure}

\begin{figure}[h]
\centering
\includegraphics[width=3in]{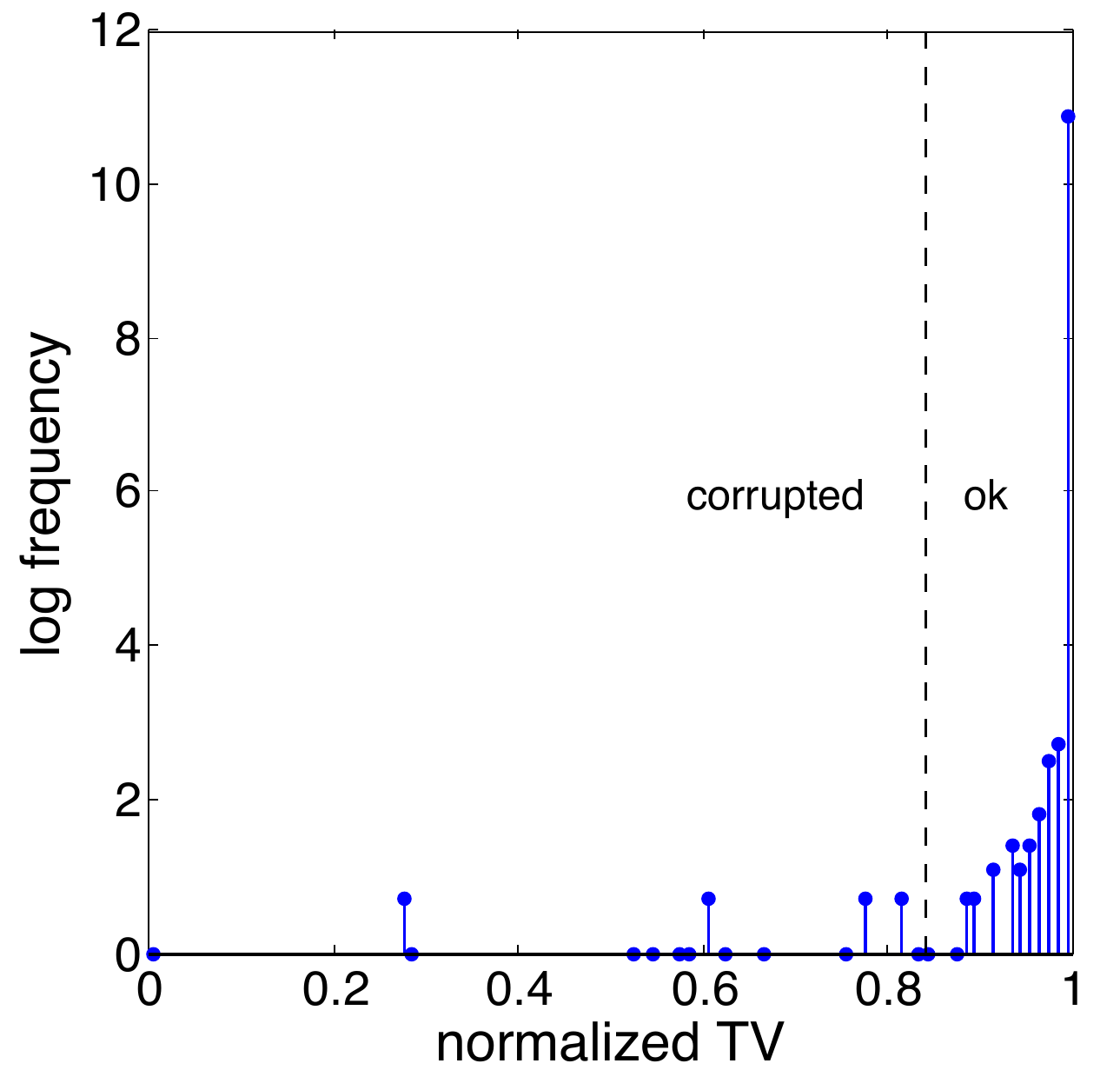}

\caption{Histogram of normalized total variation for the corrupted fat-water imaging data. The number of spikes is much smaller in this real-world case, but the general characteristics of the histogram are the same. The cutoff obtained using Otsu's method shows again a good division between the few spikes and the much more numerous non-spike-corrupted data.}
\label{realtv}
\end{figure}

\begin{figure}[h]
\centering
\includegraphics[width=\textwidth]{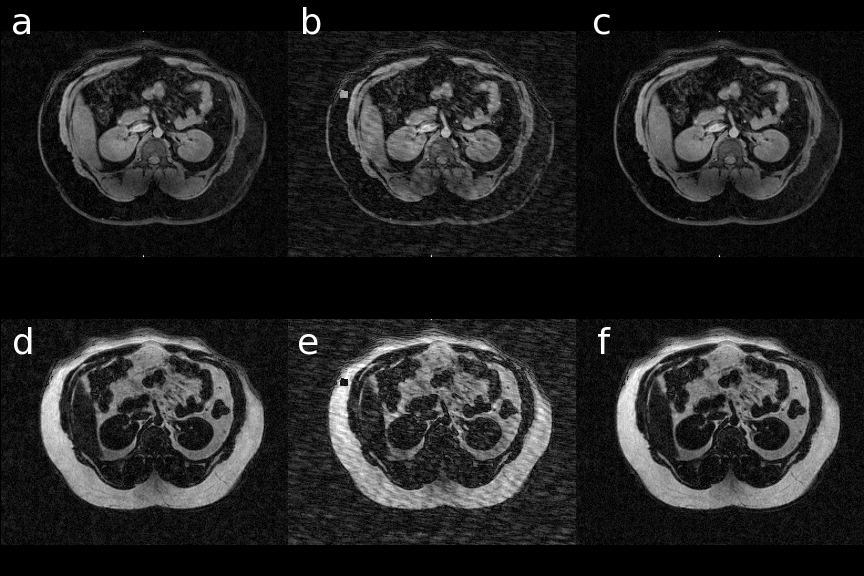}
\caption{Algorithm performance with an unknown number of spikes in
real data acquired with a faulty RF room shield during a whole-body
gradient echo scan for fat-water imaging.  The upper row (a--c)
shows the water images; the lower row (d--f) shows fat images. The
original, corrupted data is shown in the center column (b,e).  The
degradation in image quality is clear, and an area of fat
misclassification is visible in the upper left (patient right
anterior) of the images. After running this algorithm on the data,
the images are significantly cleaned, whether the corrupted data was
replaced with zeros (images a, d) or using a compressed sensing
reconstruction (c,f).  Replacing the spikes with zeros resulted in a 42\% artifact level reduction, and the CS reconstruction provided an additional 5\% reduction.}
\label{real}
\end{figure}

While the provenance of individual $k$-space points cannot be fully
known in this case, visually the data appeared to contain hundreds of
spikes spread across 240 slices.  An example slice containing spikes
is shown in Fig. \ref{realdata}. Algorithm 1 automatically detected
scores of inconsistent points and eliminated them. The resulting
reconstructed images were dramatically improved, and as a side-effect
many places throughout the derived fat-water images where water and
fat fractions were swapped were corrected. This suggests that spike
corruption can have a significant effect on derived image
quantities. The histogram of TV values produced is shown in Fig. \ref{realtv}.  With a much smaller number of spikes in this case, the form of the distribution is not as clear, but the cutoff produced by the algorithm seems to separate the relatively few number of corrupt points from the much more numerous valid points. The resulting images are shown in Fig. \ref{real}. Not only
were corrupted points correctly located, but they were replaced with
enough accuracy to improve the separation of water and fat in the
imaging data.

We measured the artifact level reduction as the mean image magnitude in the region outside of the body.  When the spikes were replaced with zeros, we found a 42\% decrease in the artifact level outside the body. When a CS reconstruction was used to replace the points, we found a 47\% decrease in artifact level.

\section{Conclusion}

We have shown a method for detecting and eliminating radio frequency
spikes in MRI data that is automated, robust, and extremely
accurate. We demonstrated the success of the technique on both 
synthetic corrupted data with known spikes and on 
whole-body gradient echo data corrupted by real RF spike noise.  Our method
achieved markedly improved images in both cases, and in the \emph{in
  vivo} experiment also improved the derived fat and water images.

The most significant contribution of Algorithm 1 is the ability to
detect corrupted $k$-space locations without human intervention.  Whatever the source of error, $k$-space data
that is inconsistent should produce a large effect on the image total
variation, a feature that can be exploited to clean images by filtering out corrupted $k$-space data.  Humans are excellent pattern recognizers, but
requiring human intervention is slow and expensive.  While
Fourier coefficient thresholding certainly works in many cases, it is easy to show an
example in which the spikes do not cross the threshold or where the
assumptions about the falloff of complex coefficients at high spatial
frequencies is incorrect.  In fact, the real spike experiment
shown here was made possible only because the spiking in this case was
missed by the MRI scanner's commercial built-in algorithm that uses Fourier amplitude thresholding.

The second step of the algorithm, to reconstruct the corrupted data
with compressed sensing, is actually only a small (11\% relative) improvement over
replacing the spiked data with zeroes, but it demonstrates the ability
of CS to accurately estimate missing data. The use of CS is justified here because the points deleted were chosen because of their effect on the TV, and CS replaces data such that the TV effect is minimized. Given the small effect of CS, though, we stress that detection and replacement of spikes, even with zeroes, is more important than implementing the ability to reconstruct the missing data with CS. We choose to show the CS results here for completeness.

Our algorithm does not blur or excessively smooth the image, as can
occur with TV-based denoising.  Since we are only undersampling the
image by a very small amount in $k$-space, the data constraint is
enforced in the Fourier domain, and we are not affecting the image as much
as image-domain techniques like TV denoising.  In fact, the amount of
data altered here is so slight that only the sinusoidal ripples were
removed, and even the background Rician noise in the magnitude image
was left intact.

We note that the artifacts due to spiking are coherent and structured,
but CS still works perfectly. The literature of compressed sensing MRI
frequently refers to the requirement that image artifacts due to the
random undersampling be ``incoherent,'' but this is somewhat
incorrect.  CS theory states that the measurement basis and the sparse
basis should be incoherent, in this case the image gradient and the
Fourier measurement.  This allows the sparse basis to maximally
constrain the data in the measurement domain, effectively turning a
global effect into information about single $k$-space points. In fact,
in the case of RF spike noise with one spike, the image artifact will
be extremely coherent, consisting of just one frequency, but TV
minimization works extraordinarily well to eliminate it because a
single errant point in $k$-space produces a global effect on the gradient
image.

While Algorithm 1 appears to be robust and flexible, many improvements
could be added to increase its accuracy and sensitivity.  For example,
inconsistencies across multiple receive coil channels or dynamic
acquisitions could supplement detection criteria to increase accuracy,
perhaps by comparing the variance of $k$-space points across time or
coil. Most functional MRI and diffusion tensor imaging studies acquire
dozens or hundreds of sequential dynamic images with echo-planar
readouts that are demanding of gradient hardware and that can
frequently cause spiking.  The advantage of the large number of
dynamics acquired in these studies is that each $k$-space location is
acquired many times, allowing temporal correlations and statistical
methods to be added to the basic TV method presented here.  Presumably, the spike distribution would be broader with a higher mean in the histogram of temporal variances for each point. Otsu's method could determine a separation threshold based on this criteria instead of the TV value alone, or perhaps both could be included in the categorization criteria.

Also, contiguous data samples instead of single $k$-space points could be deleted.  Because
spike occurrence is a temporal phenomenon and $k$-space is traversed
through time, portions of $k$-space larger than a single point may be
corrupted by a single RF event.  The data in Fig. \ref{realdata} show an example of this: the bright spikes are clustered and trail along the vertical (readout) axis. In the future, this could possibly be exploited to improve detection accuracy.


If $k$-space locations were
considered in pairs or larger groups, as well as in isolation, the
detection might be improved.  The computational penalty for this would
be large, however, since each comparison requires a full 2D or 3D fast
Fourier transform.  Note that examples shown here are 2D only, and
moving to a full 3D treatment would be significantly more
computationally intensive, but not impossible.

With different data corruption models, this approach could even be
extended to motion compensation, Nyquist ghosting, pulsatile
artifacts, etc., in which entire readout lines are corrupted or have a
systematic error.  In these cases, the image TV could constrain the
replacement of entire readout lines or help constrain empirical fits
in parametric models, such as in the case of gradient delay
estimation.
However, it is important to note that the concept of identifying inconsistent or corrupted lines of $k$-space caused by motion or other sources of data perturbation is not novel. Many methods are described in the literature to detect and compensate for artifacts caused by such k-space corruption including (but not limited to) dedicated navigator echoes  \cite{Ehman1989,Brau2006}, self-navigation \cite{Pipe1999b,Welch2004a}, iterative autofocusing \cite{Atkinson1997a,Manduca2000}, and methods based on the data redundancy afforded by multi-channel receive coils \cite{Bydder2002,Larkman2004,Atkinson2006}.

The ultimate goal of this type of approach is to create an automated
data post-processing step that detects and eliminates RF spikes and
creates a ``clean'' $k$-space for the reconstruction pipeline.  The
method presented here is compatible with a reconstruction
post-processing pipeline that works with little or no user
intervention to clean up data. Additionally, Algorithm 1 could be
modified to aid quality control checks of clinical systems and monitor
hardware function.  Finally, we have provided evidence here that with
an effective spike detection and correction method, discarding a data
set due to spiking artifacts may rarely be necessary.

\section*{Acknowledgments}

Financial support from NIBIB T32 EB001628 and NCI R25CA092043, NCATS UL1 TR000445.

\bibliographystyle{unsrt}
\bibliography{library}		

\vfill

\end{document}